# Modeling of the Output and Transfer Characteristics of Graphene Field-Effect Transistors

Brett W. Scott, *Student Member, IEEE* and Jean-Pierre Leburton, *Fellow, IEEE*

*Abstract*— We obtain the output and transfer characteristics of graphene field-effect transistors by using the charge-control model for the current, based on the solution of the Boltzmann equation in the field-dependent relaxation time approximation. Closed expressions for the conductance, transconductance and saturation voltage are derived. We found good agreement with the experimental data of Meric et al. [*Nature Nanotechnology* 3, 684 (2008)] without assuming carrier density-dependent velocity saturation.

## I. INTRODUCTION

IN recent years, graphene has emerged as a novel mono-layer material with exotic physical properties [1, 2] for applications in high performance electronic devices [3, 4]. Namely the relation between the charge carrier energy E and the two-dimensional (2D) wave vector $k = \sqrt{k_x^2 + k_y^2}$ is linear i.e. $E = \hbar v_F k$, where $v_F \sim 10^8$ cm/s is the Fermi velocity, thereby reducing the band gap to a single point (Dirac point) [1]. In this framework all carriers have a velocity with the same absolute value that is one order of magnitude larger than in conventional III-V materials [5], making graphene a promising candidate for high-speed nanoelectronics.

Recently, graphene field-effect transistors (GFETs) were successfully fabricated and exhibited *I-V* characteristics similar to conventional silicon MOS transistors [3]. Low field mobilities were however strongly degraded by the presence of coulombic space charge in the neighboring oxides, whereas

Manuscript received October 19, 2010. This work was supported by the Department of Electrical and Computer Engineering at the University of Illinois at Urbana-Champaign.
B. W. Scott is with the Department of Electrical and Computer Engineering and the Beckman Institute for Advanced Science and Technology, University of Illinois at Urbana-Champaign, Urbana, IL 61801 USA (e-mail: scott27@illinois.edu).
J. P. Leburton is with the Department of Physics, Department of Electrical and Computer Engineering, and the Beckman Institute for Advanced Science and Technology, University of Illinois at Urbana-Champaign, Urbana, IL 61801 USA (phone: 217-333-6813; e-mail: jleburto@illinois.edu).

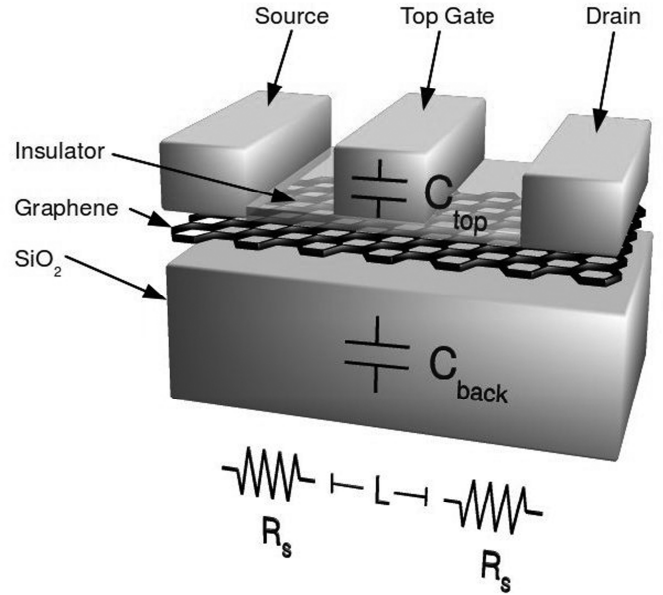

Fig. 1. Schematic of the GFET device.

nonlinearities in the current-voltage characteristics were interpreted as caused by carrier velocity saturation for which the value would depend on the carrier concentration induced by gate voltages in the 2D graphene mono-layer.

In this paper, we developed a charge-control model for GFETs that does not require the assumption of carrier density-dependent saturation velocity to reproduce the experimental characteristics. Our model also provides closed form analytic expressions for the saturation voltage, conductance and transconductance of the device.

## II. TRANSISTOR STRUCTURE

Fig. 1 shows a schematic of the GFET, where the graphene mono-layer sits on a thick $SiO_2$ layer with capacitance $C_{back}$ on top of a back gate that controls the source and drain resistance $R_s$ at the same time as the channel threshold voltage with bias $V_{gback}$. A top gate of length *L*, separated from the graphene mono-layer by a thinner oxide with capacitance $C_{top}$, controls the carriers in the channel with $V_{gtop}$. For the sake of comparison with experiment, we will only consider p-channel device operation, but our model is valid for n-channel operation as well.

### III. MATH

In order to model the transport characteristics of the GFET, we split the carrier distribution function into its even and odd parts so $f(\vec{k}) = f_{even}(\vec{k}) + f_{odd}(\vec{k})$. Then, it is well known that in the presence of randomizing collisions, and even in high fields, the Boltzmann transport equation can be written as [6]

$$\frac{eF}{\hbar} \frac{\partial}{\partial k_x} f_{even}(\vec{k}) = -\frac{1}{\tau_{tot}(k)} f_{odd}(\vec{k}) \quad (1)$$

with $\frac{1}{\tau_{tot}(k)} = \sum_i \frac{1}{\tau_i(k)}$ and the i-index indicates a particular scattering mechanism. $F$ is the electric field. In the presence of strong inter-carrier scattering for high carrier concentration, the even part of the distribution is thermalized at an electronic temperature $T_e$, and reads

$$f_{even}(\vec{k}) = \frac{1}{1 + \exp\left[\frac{\hbar v_F (k - k_F)}{k_B T_e}\right]} \quad (2)$$

where $k_F = k_F(x)$ defines the carrier concentration along the channel. In p-channel, the current can be calculated as

$$\vec{I} = \frac{4e}{L} \sum_{\vec{k}} \vec{v}(\vec{k}) f_{odd}(\vec{k}) \quad (3)$$

where $L$ is the channel length, and the factor 4 accounts for the spin and the two-fold degeneracy of the Dirac point [1]. Here $v(\vec{k}) = v_F(\cos\theta, \sin\theta)$ and $\theta$ is the angle between the electric field and the vector $\vec{k}$. Then for $\hbar v_F k_F \gg k_B T_e$ one can approximate $\frac{\partial}{\partial k_x} f_{even}(\vec{k})$ by a delta function centered around $(k - k_F)$. After integration, and given $k_F = \sqrt{\pi p}$ [7], the hole current in a 2D graphene layer reads

$$I = W \frac{2e^2}{h} F v_F \tau(p) \sqrt{\pi p} \quad (4)$$

where $W$ is the graphene layer width, $p$ is the hole concentration and $\tau(p)$ is the relaxation time for a particular carrier concentration $p$. In the high field regime, we assume $\tau(p) = \tau_{lf}/(1 + F/F_c)$ where $F_c$ is the critical field for the onset of high energy collisions such as remote phonons [8], for instance, $\tau_{lf}(p) = \tau_0 \sqrt{p/N_i}$ is the low field relaxation time dominated by scattering with charged impurities with density $N_i$ [7], and $\tau_0$ is a time constant. By setting $\mu_0 = (e/\hbar) v_F \tau_0 \sqrt{p/N_i}$, one recovers the conventional current expression

$$I = Wepv(F) \quad (5)$$

with $v(F) = \mu_0 F/(1 + F/F_c)$, where the low field conductance $\sigma_{lf} \propto p$, as observed experimentally [1, 7].

In the charge-control model, close to the Dirac point, one can use the mass action law [9] to get

$$p(x) = \frac{Q(x)}{2e} + \sqrt{\left(\frac{Q(x)}{2e}\right)^2 + p_0^2} \quad (6)$$

where $p_0$ is the minimum sheet carrier concentration [7] and $Q(x)$ is the electric charge density along the channel from source to drain given by $Q(x) = -C_{top}[V_{g0} - V(x)]$ in the gradual channel approximation [10]. Here $V_{g0} = V_{gtop} - V_0$ where $V_0$ is the threshold voltage of the GFET and is defined as [3]

$$V_0 = V_{gtop}^0 + \frac{C_{back}}{C_{top}}\left(V_{gback}^0 - V_{gback}\right) \quad (7)$$

where $V_{gtop}^0$ and $V_{gback}^0$ designate the top and back gate voltages at the Dirac point respectively. However for $\frac{Q(x)}{2e} \gg p_0$, which is the case for all bias conditions considered in this analysis, one can write

$$p(x) = \frac{Q(x)}{e}. \quad (8)$$

By integrating the current equation (5) from source to drain as in conventional MOS devices [10], and by taking into account the series resistance $R_s$ at the source and drain [3], one gets

$$I_d = \frac{W\mu_0 V_c}{2LC_{top}(|V_{ds}| - 2|I_d|R_s + V_c)}\left[Q(L)^2 - Q(0)^2\right] \quad (9)$$

where $Q(L) = -C_{top}(V_{g0} - V_{ds} - |I_d|R_s)$ and $Q(0) = -C_{top}(V_{g0} + |I_d|R_s)$. Solving for $I_d$, one obtains a closed expression for the drain current

$$I_d = \frac{1}{4R_s}\left[\begin{array}{l}V_{ds} - V_c + I_0 R_s \\ + \sqrt{(V_{ds} - V_c + I_0 R_s)^2 - 4I_0 R_s V_{ds}}\end{array}\right] \quad (10)$$

where $V_{ds}$ is the drain-source voltage, $I_0 = 2(W/L)\mu_0 V_c C_{top}(V_{gtop} - V_0 - V_{ds}/2)$ and $V_c = F_c L$.

From here, the low drain-source bias conductance is readily calculated by taking the derivative of the current expression (10) with respect to $V_{ds}$ as $V_{ds}$ goes to zero. One gets

$$g_{ds}(V_{ds} \to 0) = \frac{-V_{g0}}{|2R_s V_{g0} - R_c V_c|} \quad (11)$$

where $1/R_c = (W/L)\mu_0 C_{top} V_c$, so that $R_c V_c$ is independent on $V_c$, as is the conductance at low drain bias. The low drain-source bias resistance reads

$$R_{ds} = \frac{1}{g_{ds}} = 2R_s - \frac{R_c V_c}{V_{g0}} \quad (12)$$

which establishes a linear relation between $1/g_{ds}$ and $1/V_{g0}$ with a slope given by $R_c V_c$ (inversely proportional to the mobility) and an asymptotic conductance value for large $V_{g0}$ reaching $2R_s$.

In the same context, one obtains the expression for the drain-source saturation voltage as a function of the top gate voltage $V_{g0}$ by solving for $V_{ds}$ after setting the derivative of the current (10) with respect to $V_{ds}$ equal to zero which yields

$$V_{ds(sat)} = \frac{2\gamma V_{g0}}{1+\gamma} + \frac{1-\gamma}{(1+\gamma)^2}\left[V_c - \sqrt{V_c^2 - 2(1+\gamma)V_c V_{g0}}\right] \quad (13)$$

with $\gamma = R_s/R_c$. Substituting the drain-source saturation voltage (13) into the current equation (10) enables us to obtain the expression of the saturation drain current as a function of the top gate voltage which reads

$$I_{d(sat)} = \frac{\gamma}{R_s(1+\gamma)^2}\left[\begin{array}{l}-V_c + (1+\gamma)V_{g0} \\ + \sqrt{V_c^2 - 2(1+\gamma)V_c V_{g0}}\end{array}\right]. \quad (14)$$

By taking the derivative of the saturation current with respect to the top gate voltage one derives the expression for the transconductance at saturation,

$$g_m^{sat} = \frac{1}{R_s + R_c}\left[1 - \frac{1}{\sqrt{1 - 2(1+\gamma)V_{g0}/V_c}}\right]. \quad (15)$$

Additionally, the expression for the electric potential as a function of position along the channel length can be derived from the current equation (10) and is given by

$$V(x) = V_{g0} - V_i + \sqrt{(V_{g0} - V_i - I_d R_s)^2 + \frac{2I_d x}{W\mu_0 C_{top}}} \quad (16)$$

where $V_i = I_d/(W\mu_0 F_c C_{top})$ and the source is located at $x = 0$.

## IV. RESULTS

In this section we discuss the results for the case of $V_{gback} = -40$ V ($V_0 = 2.36$ V) and of $V_{gback} = +40$ V ($V_0 = 0.64$ V). In the former, the source and drain regions of the GFET are p-type, while in the latter, they are n-type; we notice that both threshold voltages are positive, and in both cases the top gate is biased negatively to form a p-channel.

### A. $V_{gback} = -40$ V

Figs. 2 show the plots of both the low-bias conductance $g_{ds}$ as a function of the top gate voltage, and the low-bias resistance $R_{ds}$ as a function of the inverse of the top gate voltage in the device configuration investigated in [3]. In fig. 2a the solid curve is calculated from (11) with the mobility ($\mu_0 = 550$ cm$^2$/V·s) and source resistance value ($R_s = 700$ Ω) as explicitly given in [3] which gives a good agreement with the experimental data close to the minimum conductance, but underestimates the former by about 20% at high top gate bias. The dashed curve is the best fit of (11) with the experimental conductance with $\mu_0 = 600$ cm$^2$/V·s and $R_s = 500$ Ω, which indicates that the discrepancy with the previous data is essentially due to a different value of the source resistance. In fig. 2b one can see that the experimental resistance values display a linear relation with $1/V_{g0}$ in agreement with (12). While their mobilities have similar values (within 10%) both theoretical curves are shifted from one another by the different values of the source resistance $R_{ds}$.

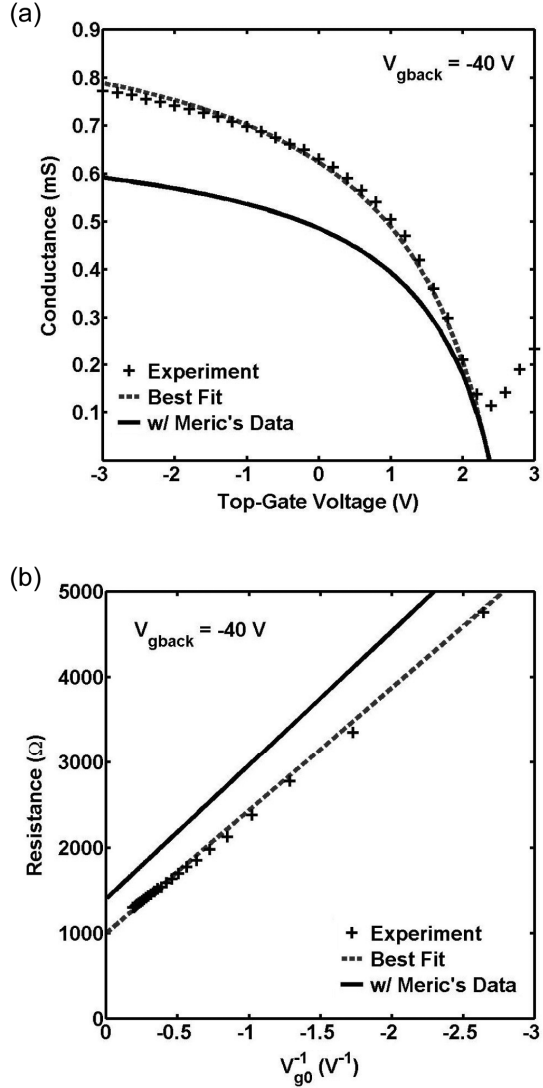

Fig. 2. (a) Small-signal source-drain conductance ($g_{ds}$) as a function of the top gate voltage minus the threshold voltage ($V_{g0}$) and (b) the small-signal source-drain resistance ($R_{ds}$) as a function of the inverse of the top gate voltage minus the threshold voltage ($1/V_{g0}$) for $V_{gback}$ = -40 V.

In fig. 3a we display the *I-V* characteristics of the GFET for $V_{gback}$ = -40 V. An excellent agreement between experiment and theory (10) is obtained with $\mu_0$ = 700 cm$^2$/V·s, $R_s$ = 800 Ω, and $V_c$ = 0.45 V for all gate biases, which provides the right current values for high (negative) $V_{ds}$. Our mobility value is 25% higher than Meric's fitted values ($\mu_0$ = 550 cm$^2$/V·s), while the source resistance is within 15% of the measured ones [3]. The up-kick in the drain current attributed to ambipolar transport for $V_{gtop}$ = 0 V is simulated by a phenomenological current term proportional to $\left(V_{ds}/V_{ds(sat)} - 1\right)^2$ [11]. For comparison, we also plot the current with the parameter values ($\mu_0$ = 600 cm$^2$/V·s and $R_s$ = 500 Ω) that best fit the conductance characteristics in fig. 2, and for which we use $V_c$ = 0.5 V ($F_c$ = 5 kV/cm) for all gate biases, which gives the right current values at $V_{ds}$ = -3 V but overestimates the current at high (negative) gate and intermediate source-drain biases. The discrepancy between the two sets of fitting parameters are within the 15-25% range, which is not really excessive and may be due to the fact that in the case of the conductance fit, the experimental data are obtained for very low bias, whereas in the case of the *I-V* fit, the mobility $\mu_0$ and source resistance $R_s$ values account for *intermediate* source-drain biases, which describe different transport processes (warm holes) with the onset of remote phonon scattering [8] at intermediate fields than low-bias transport only limited by impurity scattering [7]. Fig. 4b shows the carrier concentration (left axis) and electric potential (right axis) at the saturation onset ($V_{ds}$ = $V_{ds(sat)}$) as functions of position along the channel length with the source located at $x$ = 0. One can observe that the channel never experiences pinch-off since the carrier concentration never reaches the minimum sheet carrier concentration given by $p_0 = 0.5 \times 10^{12}$ cm$^{-2}$. Therefore we believe the current up-kick at high source drain bias for $V_{gtop}$ = 0 V may be due to other effects than electron injection from the drain side, such as impact ionization with carrier multiplication for instance [11].

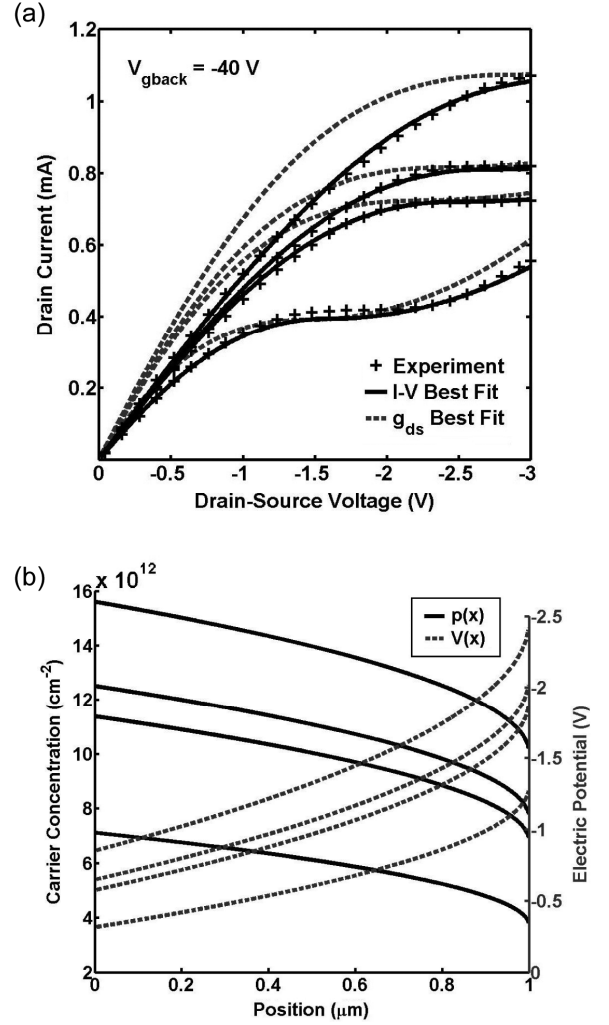

Fig. 3. (a) Drain current ($I_d$) as a function of drain-source voltage ($V_{ds}$) and (b) hole concentration (left axis) and electric potential (right axis) for $V_{ds}$ = $V_{ds(sat)}$ as functions of position along the channel length (source is on the left) for $V_{gback}$ = -40 V; $V_{gtop}$ = 0 V, -1.5 V, -1.9 V and -3 V (from bottom to top).

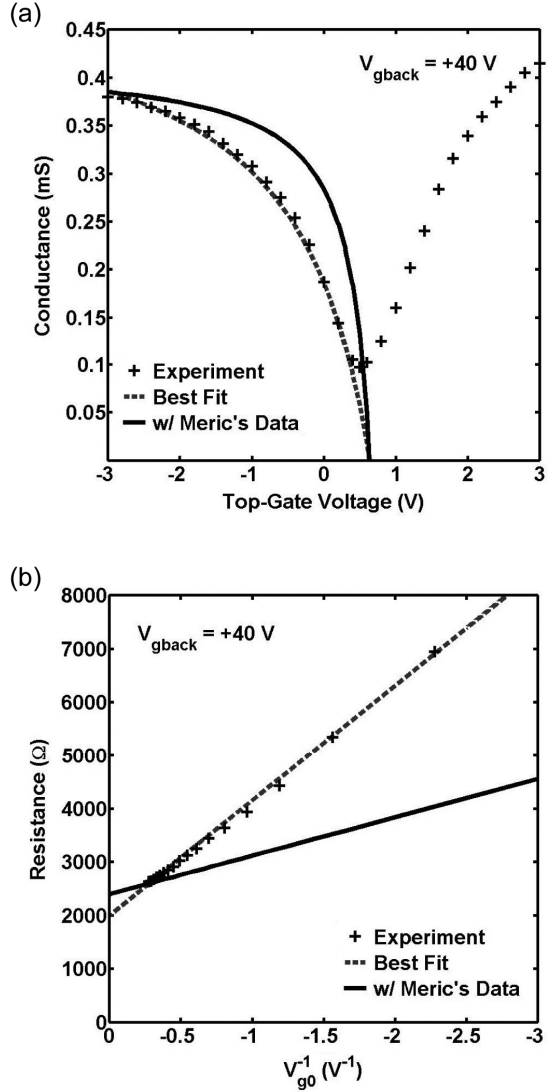

Fig. 4. (a) Small-signal source-drain conductance ($g_{ds}$) as a function of the top gate voltage minus the threshold voltage ($V_{g0}$) and (b) the small-signal source-drain resistance ($R_{ds}$) as a function of the inverse of the top gate voltage minus the threshold voltage ($1/V_{g0}$) for $V_{gback}$ = +40 V.

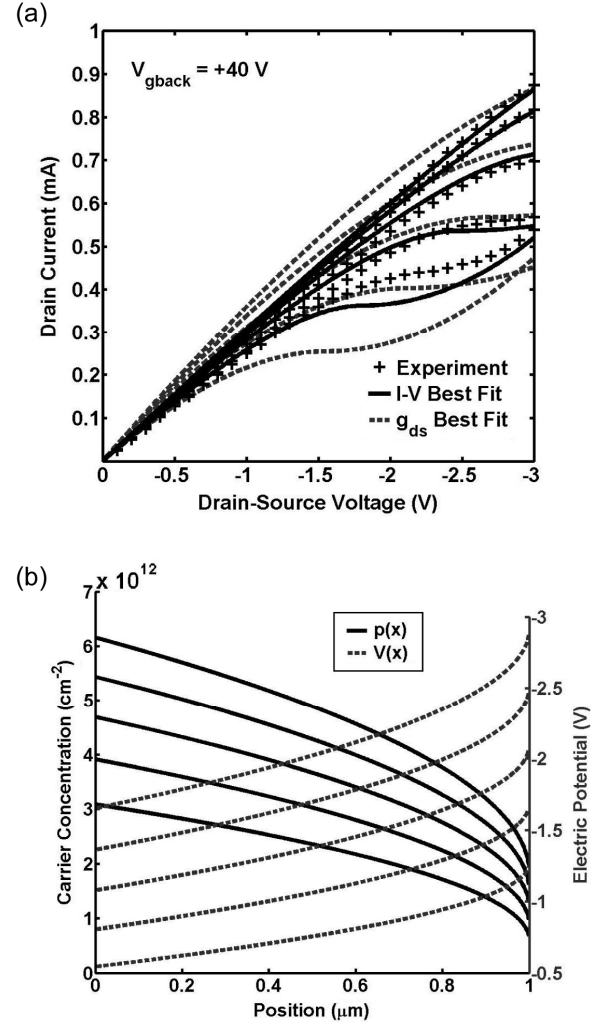

Fig. 5. (a) Drain current ($I_d$) as a function of drain-source voltage ($V_{ds}$) and (b) hole concentration (left axis) and electric potential (right axis) for $V_{ds} = V_{ds(sat)}$ as functions of position along the channel length (source is on the left) for $V_{gback}$ = +40 V; $V_{gtop}$ = -0.8 V, -1.3 V, -1.8 V, -2.3 V and -2.8 V (from bottom to top).

### B. $V_{gback}$ = +40 V

In figs. 4, we show the comparison between theoretical and experimental results for both the p-channel conductance and resistances. In fig. 4a, the solid curve is obtained from (11) with the parameters ($\mu_0 = 1200$ cm$^2$/V·s; $R_s = 1200$ Ω) given in [3], while the dashed curve uses $\mu_0 = 400$ cm$^2$/V·s and $R_s = 1000$ Ω to fit the experimental data, which again display the linear relation predicted in (12) for the resistance as seen in fig. 4b. Here, the discrepancy between the two sets of values for the fitting parameters is more dramatic since it affects both the slope (mobility), and to a less extent the asymptotic value of the source resistance.

Fig. 5a shows the I-V characteristics of the GFET for $V_{gback}$ = +40 V. Here the best fit is obtained with $\mu_0 = 1200$ cm$^2$/V·s, $R_s = 1500$ Ω and $V_c = 1.5$ V ($F_c = 15$ kV/cm) for all gate biases, which are also close to Meric's values [3], but significantly different from the best conductance fit on fig. 3 that underestimates (overestimates) the current at low (high) (negative) gate bias. This high value for $F_c$ compared to the GFET configuration with $V_{gback}$ = -40 V is indicative of the higher saturation voltage for similar channel concentrations (indeed the curves for $V_{gtop}$ = -1.8 V and -2.8 V on the one hand ($V_{gback}$ = +40 V), and $V_{gtop}$ = 0 V and -1.5 V on the other hand ($V_{gback}$ = -40 V) have similar charges at the source), while the higher source resistance provides lower current than for $V_{gback}$ = -40 V, despite the higher mobility.

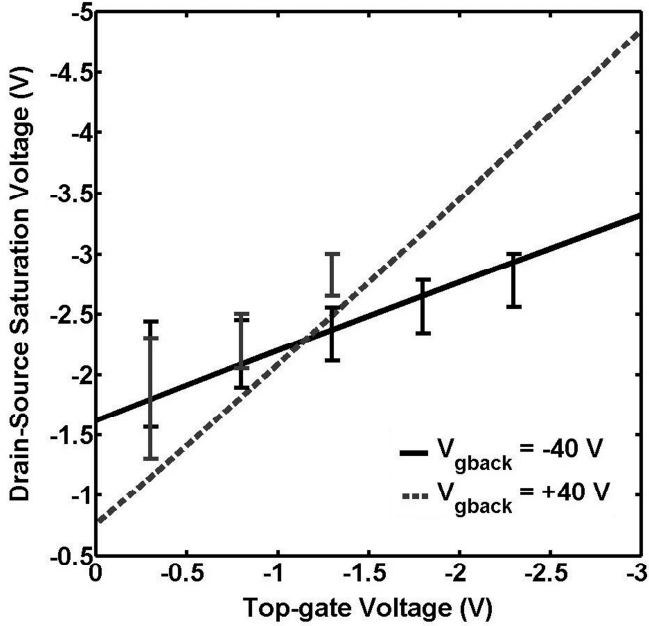

Fig. 6. Calculated drain-source voltage ($V_{ds(sat)}$) as a function of the top gate voltage ($V_{gtop}$) for two $V_{gback}$ biases. The bars are estimated values from the experimental data [3].

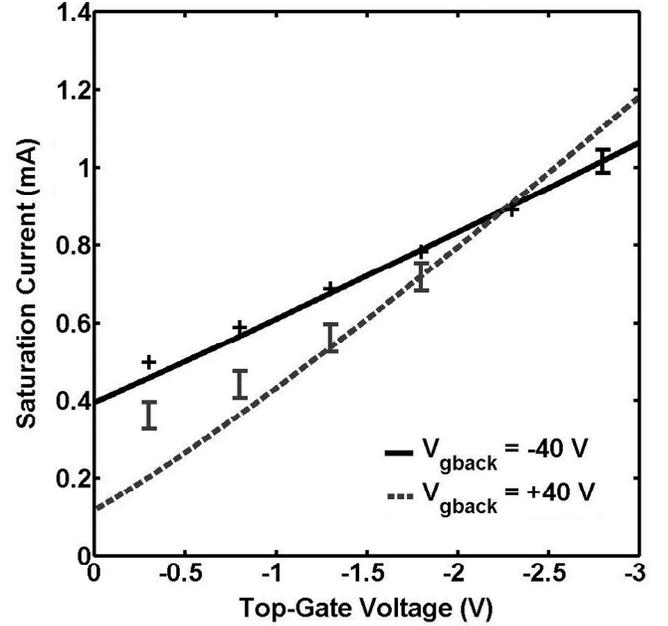

Fig. 7. Calculated saturation current ($I_{d(sat)}$) as a function of the top gate voltage ($V_{gtop}$) for two $V_{gback}$ biases. Crosses (+) and bars (I) are estimated from the experimental data [3].

Fig. 5b shows the carrier concentration and electric potential at the saturation onset ($V_{ds} = V_{ds(sat)}$) as functions of position along the channel length for the different gate biases. From a general standpoint, carrier concentrations are lower than for the case with $V_{gback}$ = -40 V, because of the lower threshold voltage ($V_0$ = 0.64 V instead of $V_0$ = 2.36 V). Again, it can be seen that the channel never experiences pinch-off since the carrier concentration never reaches the minimum sheet carrier concentration.

### C. Transfer Characteristics

In fig. 6 we plot the drain-source saturation voltage as a function of gate bias (13) for the two GFET configurations. The vertical bars on the plot represent the approximate range of the saturation drain-source voltage obtained from the experimental plots [3]. One notices the excellent agreement between theory and experiment, especially for the $V_{gback}$ = -40 V condition, whereas the discrepancy for the $V_{gback}$ = +40 V configuration is due to the uncertainty in ascertaining the experimental values that fall out of the figure. One also notices the steeper variation of the saturation voltage in the latter case compared to the former case, which is reflected in the larger value of the critical fields to reproduce the experimental data.

In fig. 7 we plot the saturation current as a function of the top gate voltage (14). For the case $V_{gback}$ = -40 V, the extraction of the experimental values of the saturation current is straightforward, except for high top gate biases for which the current has not saturated (fig. 3a), and shows an excellent agreement with our model. For the case $V_{gback}$ = +40 V the bars are estimates of experimental values because the current does not saturate for all values of $V_{gtop}$ over the range of the source-drain voltage (fig. 5a). For both $V_{gback}$, it can be seen that the relationship between the saturation current and top gate voltage is linear, and an excellent agreement between theory and experiment is obtained with discrepancies occurring at low top gate biases.

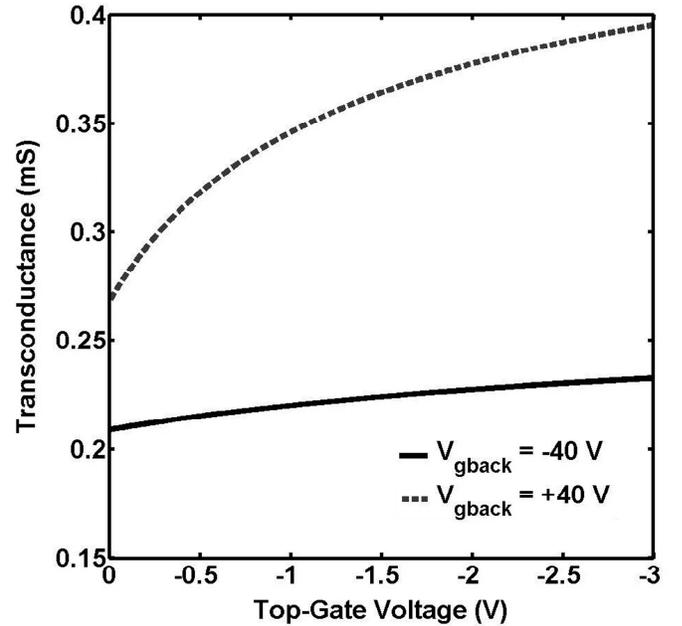

Fig. 8. Calculated transconductance at saturation ($g_m^{sat}$) as a function of the top gate voltage ($V_{gtop}$) for two $V_{gback}$ biases.

We also display the profile of the transconductance at saturation as a function of top gate voltage (15) in fig. 8. One notices that for $V_{gback}$ = +40 V, $g_m$ is much more drastically affected by the variation of the top gate voltage than for the $V_{gback}$ = -40 V condition. This is due to the $V_c$ term in (15) since the critical field, and consequently the critical voltage, is much larger when $V_{gback}$ = +40 V.

## V. Discussion and Conclusions

We provide a coherent model for the output and transfer characteristics of GFETs with two back-gate bias configurations, for which the source and drain contacts are either p- or n-type. For unipolar transport, closed form expressions are obtained for the current, low drain bias conductance, transconductance at saturation, saturation voltages, saturation currents and potential along the channel, which rely on three parameters i.e. low field carrier mobility, source-drain resistance and critical field for the high energy carrier scattering, to reproduce the experimental I-V characteristics for each back-gate condition. In particular we predict a linear dependence of the low-field resistance versus inverse gate voltage, which is quantitatively confirmed, while we point out a discrepancy between the parameter values used for the $g_{ds}$-$V_{g0}$ plots and the I-V characteristics, especially for positive back gate voltage, which has not been resolved so far. However the predicted quasi-linear dependence between saturation voltage and gate voltage is well confirmed experimentally.

Let us emphasize that our model relies on only one $F_c$ parameter to describe the current at high drain biases for all top gate biases, which according to the velocity field relation $v(F)$ implies a single saturation velocity $v_{sat}$ = 3.2 x $10^6$ cm/s (1.8 x $10^7$ cm/s) for $V_{gback}$ = -40 V (+40 V), unlike Meric's model that requires a concentration dependent saturation velocity to fit the experimental data. In this respect, let us point out that close analysis of the source-drain field profile indicates that the maximum fields achieved in the highest drain biases are only a few times the critical field values $F_c$, which is far from achieving saturation; it is therefore quite possible that the velocity-field relation acquires a lower slope due to remote phonon scattering rather than saturating [12].

Finally, we also point out that detailed analysis of the charge control model indicates that even for the lowest (negative) top gate bias i.e. $V_{gtop}$ = 0 V (-0.8 V) for $V_{gback}$ = -40 V (+40 V), the channel never reaches pinch-off, which suggests that the current increase at high drain biases may be due to other causes than electron injection [11].